\documentclass[12pt,epsf]{article}

\begin{document}
\newcommand{\gsim}{ \mathop{}_{\textstyle \sim}^{\textstyle >} }
\newcommand{\lsim}{ \mathop{}_{\textstyle \sim}^{\textstyle <} }
\newcommand{\drawsquare}[2]{\hbox{%
\rule{#2pt}{#1pt}\hskip-#2pt
\rule{#1pt}{#2pt}\hskip-#1pt
\rule[#1pt]{#1pt}{#2pt}}\rule[#1pt]{#2pt}{#2pt}\hskip-#2pt
\rule{#2pt}{#1pt}}
\newcommand{\fund}{\drawsquare{6.5}{0.4}}
\newcommand{\symm}{\drawsquare{6.5}{0.4}\hskip-0.4pt%
        \drawsquare{6.5}{0.4}}
\baselineskip 0.7cm

\begin{titlepage}

\begin{flushright}
UT-05-13
\end{flushright}

\vskip 1.35cm
\begin{center}
{\large \bf
More on Conformally Sequestered SUSY Breaking
}
\vskip 1.2cm
M.~Ibe${}^{1}$, Izawa~K.-I.${}^{1,2}$, Y.~Nakayama${}^{1}$,
Y.~Shinbara${}^{1}$, and T.~Yanagida${}^{1,2}$
\vskip 0.4cm

${}^1${\it Department of Physics, University of Tokyo,\\
     Tokyo 113-0033, Japan}

${}^2${\it Research Center for the Early Universe, University of Tokyo,\\
     Tokyo 113-0033, Japan}

\vskip 1.5cm

\abstract{
We extend our models for conformal sequestering
of dynamical supersymmetry breaking
with decoupling vector-like matter
in several different ways.
These extensions enable us to simplify concrete model building,
in particular, rendering large gauge group and {\it ad hoc} global symmetry
for sequestering unnecessary.
Conformal sequestering appears highly natural in such circumstances.
}
\end{center}
\end{titlepage}

\setcounter{page}{2}

\section{Introduction}

It is phenomenologically interesting to study 
superconformal gauge theory
(see Ref.~\cite{BZ,SC,CGT}),
since if it includes a SUSY-breaking 
sector, conformal sequestering~\cite{LS}
of the SUSY breaking may take place,
providing a solution to the flavor-changing 
neutral current (FCNC) problem in the supersymmetric standard model.%
\footnote{See also Ref.~\cite{Dine,Sundrum}.
For some other phenomenological applications
of superconformal dynamics, see Ref.~\cite{NS}.}

In a previous paper~\cite{Ibe:2005pj},
we modified vector-like gauge theories for the SUSY
breaking~\cite{IYIT} by adding massive
hyperquarks to turn the full high-energy theory above the mass threshold 
into conformal gauge theory.
More generally, our strategy
to construct conformally sequestered hidden sector
is as follows:
We first choose a model of dynamical SUSY breaking.
Next, we add vector-like supermultiplets $\Phi$ (hyperquarks) to uplift the
SUSY-breaking model to a conformal field
theory in the anticipation of introducing a mass term for $\Phi$.%
\footnote{This mass can be originated from a vacuum expectation value
of another field (see section~\ref{sec:U1Yukawa}).}
Then, the theory which starts at the Planck scale flows to the  infrared (IR) fixed 
point of the conformal theory.
Finally, the mass of $\Phi$ breaks the conformal invariance
and effectively leads to the above model of
dynamical SUSY breaking.

However, as is detailed in the previous paper, this simple modification
does not achieve the conformal sequestering due to
unwanted global $U(1)$ symmetries
accompanied by the introduction of hyperquarks $\Phi$.
In order to eliminate the unwanted global $U(1)$ symmetries,
we introduced non-abelian gauge interactions acting on the additional
massive hyperquarks.
These additional interactions are so determined as to be strong enough
to break the $U(1)$ symmetries for sufficient sequestering effects.

Although we obtained various examples realizing the sequestering,
the concrete models so constructed typically have large gauge groups
and also indispensable global non-abelian symmetry of the SUSY-breaking sector.
In particular, since the global non-abelian symmetries, which are thought to be enhanced 
at the IR fixed point, prevent a generic K{\" a}hler potential from  sequestering, 
we were forced to assume the presence of the global symmetry from the start  
in the hidden sector.

In this paper, 
we further extend our models for conformal sequestering
of dynamical SUSY breaking
with decoupling vector-like matter
in several different ways.
These extensions enable us to simplify concrete model building,
in particular, rendering large gauge group and global symmetry
for sequestering unnecessary.

The rest of the paper goes as follows:
In section~\ref{sec:sequestering}, we summarize  generic problems to achieve the 
conformal sequestering.
In section~\ref{sec:U1}, we present concrete methods to eliminate problematic $U(1)$
symmetries which disturb the conformal sequestering. 
In addition to the way of breaking $U(1)$ symmetries by introducing gauge interactions 
discussed in Ref.~\cite{Ibe:2005pj},
we pursue another way of introducing relevant deformations in the superpotential.
In section~\ref{sec:noFlavor}, we explain the models which require no global non-abelian symmetry
in the hidden sector imposed by hand.
Our models are based on a so-called non-calculable SUSY breaking model~\cite{Affleck:1984mf} 
of the $SO(10)$ gauge theory which includes only one chiral superfield
in the 16-dimensional spinor representation.
Finally, in section~\ref{sec:noSym}, we discuss a model which is defined by
adding massive vector-like matter to the SUSY breaking model in Ref.~\cite{Murayama:1995ng}.
In the construction of the final model, we need no new interactions to forbid
problematic $U(1)$ symmetries nor non-abelian symmetries imposed by hand, 
and hence, the model is natural in the sense of Ref.~\cite{Dine}.

\section{Conformal sequestering}\label{sec:sequestering}

The K{\" a}hler potential interaction between hidden sector superfields $a_i$ and
visible sector ones $q_a$
\begin{eqnarray}
{\mit \Delta} K = \frac{C^{ab}_{ij}}{M_P^2}q^\dagger_aq_b
a^\dagger_i a_j
 \label{kahler}
\end{eqnarray}
induces a severe FCNC problem for generic $C^{ab}_{ij}$.
The conformal sequestering \cite{LS} is intended to achieve small
$C^{ab}_{ij}$ at low energy
by means of manageable strong dynamics of the hidden sector.
 
Let us suppose that the hidden sector flows to a strongly coupled
superconformal field theory (SCFT) through a certain high-energy scale 
$\Lambda_{CFT}$ to a small mass scale $m$ which eventually sets the SUSY breaking scale.
Owing to the large renormalization effects of the SCFT,
 $C^{ab}_{ij}$ at low energy are expected to be suppressed (that is, the hidden sector sequestered) as
\begin{eqnarray}
C^{ab}_{ij} (m)\propto \sum_k C^{ab}_{k}e^{-L_{ijk}\ln {\Lambda_{CFT} \over m}},
\end{eqnarray}
The matrix $L_{ijk}$ is to be called as a sequestering matrix,
whose component values are obtained from the anomalous dimensions%
\footnote{The operator mixing makes the anomalous dimensions a matrix
indexed by ${ij}$ and $k$.}
of the (possibly non-conserved) 
composite current superfield $[a_ia_j^\dagger]_r$.
The determinant of $L_{ijk}$ is vanishing for conserved currents of the SCFT,
because the conserved current is not renormalized.
The K{\" a}hler term corresponding to a zero eigenvalue is not suppressed
and then the sequestering is not achieved. 

Our primary concern is  on the $U(1)$ part of the sequestering matrix,
that is, the $i=j$ part $L_{ik}=L_{iik}$.
As for this part, the sequestering matrix is related to
the slopes
of the $\beta$ functions with respect to
the coupling constants $g_k$ in the SCFT
as $L_{ik} \sim \partial_{g_k}\beta_i|_{*}$,
or the slopes of the  anomalous dimensions
$\gamma_{a_i} = -(\partial\ln Z_{a_i}/{\partial t})$
of the elementary fields
as $L_{ik} \sim \partial_{g_k} \gamma_{a_i}|_* $,%
\footnote{The relation between the $\beta$ function and the anomalous
dimension of the elementary field is given by the NSVZ formula
\cite{NSVZ}.
The slope of the $\beta$ function is related to the anomalous
dimension of the composite current operator as {\it e.g.} in
Ref.~\cite{Anselmi:1996mq}.}
where ``$*$" indicates the values evaluated at the fixed point
and $t=\ln(\mu/\Lambda_{CFT})$ with $\mu$ as the renormalization scale.%
\footnote{See Refs.~\cite{LS,Ibe:2005pj} for details.
We have shown that
the sequestering matrix for the $U(1)$ part is
given by the Hessian of the renormalization group flow at the fixed
point when there is no $U(1)$ symmetry \cite{Ibe:2005pj}.
}

Therefore, to construct a conformally sequestered SUSY-breaking model,
we should look for a SCFT with no (abelian) conserved currents.
As is explained in the Introduction,
we begin with a dynamical SUSY-breaking model 
and add several vector-like matters to turn it into a SCFT.
Typically, the very introduction of additional vector-like matters $\Phi$%
results in an enhancement of $U(1)$ symmetries.\footnote{The small mass terms $m \ll \Lambda_{CFT}$ for these
vector-like matters may be ignored in discussing the superconformal
dynamics relevant for sequestering.
}
They stem from non-anomalous combinations of the $U(1)$ axial rotations
on the $\Phi$ fields
and anomalous $U(1)$ axial rotations in the SUSY-breaking part.

In the next section, we discuss a few ways
to eliminate such $U(1)$ symmetries by introducing further interactions.
On the other hand, in section~\ref{sec:noSym},
we present a SUSY breaking model
where the introduction of a vector-like field does not result
in an enhancement of  $U(1)$ symmetry, and hence, no additional
interactions are  required.

Finally, we consider the non-abelian part, namely $i\neq j$. 
As for the non-abelian part, we can forbid the corresponding
K{\" a}hler couplings Eq.~(\ref{kahler}) by imposing a global symmetry on the hidden sector,  
that is, we can take $C_{ij}^{ab} \propto C^{ab}\delta_{ij}$
naturally in the sense of 't Hooft.
As such, the problem is reduced to the existence of  $U(1)$ symmetries as discussed above.
However, imposing such a global non-abelian symmetry seems
``unnatural" as emphasized in Ref.~\cite{Dine}.

The distinguishing feature, which solves this ``naturalness" problem, of our models presented in sections~\ref{sec:noFlavor} and \ref{sec:noSym}
is that they do not require
any global non-abelian symmetries imposed on the hidden sector
by judiciously choosing the matter contents and interactions.

\section{Eliminating unwanted $U(1)$ symmetries}\label{sec:U1}

In this section, we consider a few possibilities to break those
unwanted $U(1)$ symmetries with conformality of the dynamics kept intact.
The breaking effects should be
large because the amount of sequestering follows that of the breaking.
This condition implies necessity to construct another strongly
coupled SCFT through relevant deformations of
the above SCFT with the $U(1)$ symmetries.
At the same time, we should arrange the total model so
that SUSY breaking is also realized in the end.
Namely, a possible recovery of SUSY
through the deformations should be avoided.%
\footnote{Although SUSY-broken vacua may exist as local ones
in such SUSY recovery examples,
it seems hard to separate a SUSY-broken vacuum and a supersymmetric
one far enough to stabilize the former,
since the deformations need to be sizable for sufficient
conformal sequestering.}

\subsection{by gauge  interactions}\label{sec:U1gauge}

One way to break an additional $U(1)$ symmetry (rotation of $\Phi$) is to
introduce additional gauge interaction
on the matter field $\Phi$. When the $U(1)$ symmetry is chiral under the additional gauge interaction, the $U(1)$ symmetry is broken via
the associated chiral anomaly. This possibility was pursued in
Ref.~\cite{Ibe:2005pj}. A typical example is the $SP(3)\times SP(1)^2$ gauge
theory with matter contents given
in table~\ref{tab:content2}. The low-energy SUSY breaking
is provided by the $SP(3)$ IYIT model
\cite{IYIT},
and additional quark superfields $Q'$ (as $\Phi$)
are gauged under $SP(1) \times SP(1)$ to break $U(1)$ symmetry that
rotates $S_{ij}$, $Q^i$, and $Q'$ simultaneously. The superpotential
has a form
\begin{eqnarray}
W = \lambda S_{ij} Q^{i}Q^{j} + m Q'^2 \ ,
\end{eqnarray}
where the contracted gauge indices are omitted.

\begin{table}[tb]
\begin{center}
\begin{tabular}{c|ccc}
        &$SP(N)$&$SP(N')$&$SP(N')$\\
 \hline
$2(N+1) \times Q$ & $\fund_{2N} $ & $\bf 1$ & $\bf 1$\\
\hline
$Q'$ & $\fund_{2N} $ & $\fund_{2N'}$ & $\bf 1$\\
$Q'$ & $\fund_{2N} $ & $\bf 1$ & $\fund_{2N'}$\\
\hline
$S_{ij}$ & $\bf 1 $ & $\bf 1$ & $\bf 1$ \\
\end{tabular}
\end{center}
\caption{The matter contents in the strongly coupled
$SP(N)\times SP(N')^2$ model with $N=3, N'=1$.
Here, the subscripts of the fundamental representations
denote the dimensions of the representations.
In terms of the $SP(N)$ gauge theory,
the number of the fundamental representation is given by
$N_F = 2 (N+1) + 2\times 2 N'=12$, 
while the number of the fundamental representations
of each $SP(N')$ gauge theory is given by
$N_F' = 2N = 6$.}
\label{tab:content2}
\end{table}

For $m=0$, the theory is expected to have
a strongly-coupled IR fixed point for
the gauge couplings $g_{{}_{SP(3)}}$,
$g_{{}_{SP(1)\times SP(1)}}$ and Yukawa coupling
$\lambda$. The slopes of the anomalous dimensions ({\it i.e.} sequestering
matrix) are expected to be of order one though the explicit
computation is hard to perform. We note that the structure of the
gauged SCFT (anomalous dimension, central charge, {\it etc.}) is totally
different from the original ungauged SCFT.
As for the SUSY breaking, the mass term for $Q'$
in the superpotential exclusively causes no problem.

\subsection{by Yukawa interactions}\label{sec:U1Yukawa}

Another possible way to break the $U(1)$ symmetries is to add 
superpotential terms (Yukawa interactions) that are relevant deformations 
of the SCFT, which lead to a new strongly coupled CFT and break the unwanted symmetries explicitly.

Let us introduce a singlet $Y$ and
try a superpotential
\begin{eqnarray}
W = \lambda S_{ij} Q^{i}Q^{j} + \lambda_Y Y \Phi^2 + \lambda_n Y^n,
\end{eqnarray}
where $\lambda_{Y}$ and $\lambda_{n}$ denote the coupling constants and 
$3/2 \le n \le 3$ for unitarity at a possible fixed point.
This deformation completely eliminates the axial symmetry of $\Phi$ rotation
(except a $U(1)_R$ symmetry).
If the deformation is relevant, that is, a choice of the coupling constants
yields a nontrivial fixed point, it provides a candidate SCFT for
conformal sequestering.

However, the simple addition of the mass term $m\Phi^2$ does not
bring it back to the SUSY-breaking model as discussed in Appendix A.
To retain broken SUSY, we further introduce another singlet $Z$
and add superpotential terms
\begin{eqnarray}
{\mit \Delta}W = MZ(Y-m),
\end{eqnarray}
where $M$ denotes a mass scale near $\Lambda_{CFT}$.

Under this deformation, the conformal sequestering and the subsequent
dynamical SUSY breaking go as follows:
The introduction of the singlet $Z$ without the superpotential
would be accompanied by a new conserved $U(1)$ current
due to the rotation of $Z$.
However, we expect that the interaction $MZY$ will lead to a new
CFT point, breaking this $U(1)$ symmetry and realizing conformal
sequestering of all the hidden fields appearing in the action.
The tadpole term
$-mM Z$ then serves as a relevant deformation, which cannot have a nontrivial fixed point, and eventually yields a mass of order $\lambda m$ to
$\Phi$. Once the $\Phi$ field becomes massive, the low-energy dynamics is
described by the IYIT model and the dynamical SUSY breaking at low
energy is achieved.%
\footnote{
This Yukawa-type deformation
possibly has an intimate relationship with the gauge-type one.
An example to indicate this is given in Appendix B.}

\section{Models with no imposed flavor symmetry}\label{sec:noFlavor}

Now that we have eliminated the problematic $U(1)$ symmetries
of the SCFT, we turn to consider the non-abelian symmetry thereof.
For example, the models that exemplified sequestering
in the previous section require 
the $SU(4)$ symmetry for the hidden sector
\cite{Ibe:2005pj}.
The ``naturalness'' problem of such non-abelian symmetry for conformal
sequestering is emphasized
in Ref.~\cite{Dine}.
In this section, we provide a concrete example of conformally sequestered
SUSY-breaking model with no global symmetry (except $U(1)_R$)
in the hidden sector.

Let us construct a conformally sequestered SUSY-breaking model with no
global symmetry step by step, following the procedure exposed in the
preceding sections.

\newpage

\begin{itemize}
\item{\bf SUSY-breaking sector}
\end{itemize}
SUSY-breaking sector is provided by a so-called
non-calculable model~\cite{Affleck:1984mf} of the $SO(10)$ gauge theory
with one chiral superfield $\psi$
in the 16-dimensional spinor representation.
Non-calculable models have no classical flat direction and consequently the
SUSY is expected to be broken dynamically:
the 't~Hooft anomaly-matching
condition seems difficult to be satisfied
when we have no calculable
description of the low-energy dynamics at the classical level.
The above model is unique in the point that
the chiral content for SUSY-breaking consists of a single multiplet.%
\footnote{As far as we know, the supersymmetric $SU(2)$ gauge theory
with a chiral multiplet of the five-dimensional representation~\cite{ISS}
gives the other candidate with this property. Unfortunately,
we do not have a conformal extension of that example.}
\begin{itemize}
\item{\bf Conformal extension}
\end{itemize}
In order to attain the conformal sequestering, we add vector-like
multiplets to make the model conformal. There are several choices of
additional matter multiplets. One possibility is to add many ${\bf 10}$'s
($H_i$ for $i = 1,\cdots, N_f$).
Then the theory
(without superpotential terms) is expected to flow to a nontrivial
conformal fixed point for $7\le N_f \le 21$~\cite{Pouliot:1996zh}.%
\footnote{As can be seen from table \ref{table1},
$a$-maximization implies
an enhanced $U(1)$ symmetry
(see Ref.~\cite{Ans})
in the IR fixed point for $7\le N_f \le 9$,
which should remedy the unitarity-violating charge assignment.
This class of models has been investigated recently \cite{private}.
Hereafter we restrict ourselves to $N_f \ge 10$.}
The $\beta$ function for
the gauge coupling $\alpha = {g^2}/{4\pi}$ is given
by
\begin{eqnarray}
\beta_{\alpha} = -\alpha^2 \frac{3\times (10-2)
 - N_f(1-\gamma_{10}) -2 (1-\gamma_{16})}{2\pi-8\alpha} \ , \label{nsvz}
\end{eqnarray}
where $\gamma_r$ is the anomalous dimension of the chiral superfield in
the $r$-dimensional representation.

The model has $SU(N_f)\times U(1)$ symmetry along with the
conformal $U(1)_R$ symmetry, whose charge $R$ of a chiral operator with naive dimension one 
is related to its anomalous dimension $\gamma$
at the fixed point by the formula $R = \frac{2}{3}(1+\frac{\gamma}{2})$
\cite{SC}.
The $R$ charges can be obtained by the
$a$-maximization procedure \cite{Intriligator:2003jj}:
through maximizing the $a$-function
\begin{eqnarray}
a(R) &=& \sum [3(R-1)^3 - (R-1)] \nonumber \\
 &=& 16[3(R_{16}-1)^3 - (R_{16}-1)] + 10N_f[3(R_{10}-1)^3 - (R_{10}-1)]
\end{eqnarray}
under the condition that the $\beta$ function Eq.(\ref{nsvz}) vanishes,
we obtain
\begin{eqnarray}
R_{10} = \frac{-15-24N_f + 3N_f^2 +\sqrt{2885-N_f^2}}{3(-5+N_f^2)}.
\end{eqnarray}
The numerical results are summarized in table~\ref{table1}.
The above conformal extensions will reduce to the SUSY-breaking
model after introducing the mass terms for the $\bf 10$'s.

\begin{table}[bp]
\begin{center}
\begin{tabular}{|l|r|l|r|} \hline
 & $R_{10}$ & $\quad R_{16}$ & $R_{54}$ \\\hline
$N_{10}=7, N_{16}=1$ & 0.131 & 0.0415 & --- \\\hline
$N_{10}=8, N_{16}=1$ & 0.215 & 0.140 &  --- \\\hline
$N_{10}=9, N_{16}=1$ & 0.285 & 0.2175 & --- \\\hline
$N_{10}=10, N_{16}=1$ & 0.343 & 0.285 & --- \\\hline
$N_{10}=14, N_{16}=1$ & 0.504 & 0.472 & --- \\\hline
$N_{10}=16, N_{16}=1$ & 0.558 & 0.536 & --- \\\hline
$N_{10}=1, N_{16}=1, N_{54}=1$ & 0.514 & 0.483 & 0.377 \\\hline
\end{tabular}
\end{center}
\caption{The $R$ charge assignments which are determined from the $a$-maximization
procedure.
$N_r$ denotes the number of the chiral superfields in the $r$-dimensional representation.
}
\label{table1}
\end{table}

\begin{itemize}
\item{\bf Breaking of unwanted $U(1)$}
\end{itemize}
The additional matter causes
new conserved $U(1)$ currents.
We should eliminate the unwanted $U(1)$ symmetries so that the conformal
sequestering occurs. 

{\bf i) gauge couplings }\\
As discussed in section~\ref{sec:U1}, one way to do this is to gauge the
flavor symmetry, which leads to the anomalous breaking of the $U(1)$
symmetry.
Let us show how it works in our present setup by considering $SO(10)$
gauge theory with fourteen ${\bf 10}$'s and sixteen ${\bf 10}$'s
in turn.

\begin{table}[tb]
\begin{center}
\begin{tabular}{c|ccc}
        &$SO(10)$&$SO({N_f}/{2})$&$SO({N_f}/{2})$\\
 \hline
$\psi$ & spinor({\bf 16}) & $\bf 1$ & $\bf 1$\\
\hline
$H_1$ & $\fund_{10} $ & $\fund_{{N_f}/{2}}$ & $\bf 1$\\
$H_2$ & $\fund_{10} $ & $\bf 1$ & $\fund_{{N_f}/{2}}$\\
\hline
$S$ & $\symm_{54} $ & $\bf 1$ & $\bf 1$ \\
\end{tabular}
\end{center}
\caption{The matter contents of the $SO(10) \times SO({N_f}/{2}) \times SO({N_f}/{2})$ model with $\psi$ and $H$'s. The symmetric traceless representation $S({\bf 54})$ is written here for later use.}
\label{tab:content1}
\end{table}

For example, we gauge the flavor symmetry by $SO(7)\times SO(7)$ gauge group 
with the matter contents summarized in table~\ref{tab:content1}.%
\footnote{
As a gauged flavor symmetry,
$SP$ would forbid the mass term for $10$'s and $SU$ would result
in unwanted vector-like $U(1)$ symmetries.}
Now we argue that the resultant theory flows to a new conformal theory in
the IR. The perturbation by 
$SO(7) \times SO(7)$ interaction yields a relevant deformation
of the original SCFT, which is expected to
flow to a new fixed point in the IR. The
conformal $R$ charges of this gauged theory are obtained as
\begin{eqnarray}
R_{10} = R_{16} = \frac{1}{2} 
\end{eqnarray}
corresponding to $\gamma_{10} = \gamma_{16} = -\frac{1}{2}$.

The sequestering matrix near the fixed point is
determined by first-order deferential equations for ${\mit \Delta}
\ln Z_i \equiv \ln Z_i + \gamma_i^*t$ because the FCNC causing
interaction in Eq.~(\ref{kahler}) is regarded as the initial value of
${\mit \Delta}\ln Z_i \supset
\frac{c_i^{ab}}{M_P^2}q_a^\dagger q_b$. Under the renormalization
convention given in Ref.~\cite{LS}, they are given by
\begin{eqnarray}
\frac{d}{dt} {\mit \Delta}\ln Z_i = 
- \sum_a\left(\frac{\partial\gamma_i}{\partial\alpha_a}\right)\bigg|_*
{\mit \Delta}\alpha_{a} = \sum_kL_{ik} {\mit \Delta}\ln Z_k ,
\end{eqnarray}
where 
\begin{eqnarray}
{\mit \Delta}\alpha_{SO(10)}  &=&
\frac{\alpha_{SO(10)}^{*2}}{2\pi - 8\alpha^*_{SO(10)}}[N_f{\mit \Delta}\ln Z_{10}
 + 2{\mit \Delta}\ln Z_{16}], \nonumber\\
{\mit \Delta}\alpha_{SO(7)}  &=&
\frac{\alpha_{SO(7)}^{*2}}{2\pi - 5\alpha^*_{SO(7)}}[{N'_f}{\mit \Delta}\ln Z_{10}],
\end{eqnarray}
with $N_f=14$ and $N_f' = 10$ in this particular model. 
We expect that the sequestering matrix is of the same
order in magnitude as the anomalous dimensions of order one to induce
sufficient sequestering.%
\footnote{However, there is one subtlety
here. Although the added gauge interaction becomes strong in the IR, the
new conformal fixed point might be too close to the ungauged conformal fixed
point as could be inferred from the conformal $R$ charge
assignment in table 2. Thus, the conformal sequestering might
be insufficient in this
case. Owing to this possibility, the next example seems preferable.}

Another example is the $SO(8)\times SO(8)$ gauge interaction
with $R_{10} = \frac{2}{5}$,
$R_{16} = \frac{9}{5}$. The anomalous dimensions are relatively large
and considerably away from the original fixed point
with a $U(1)$ conserved current. We expect that this model yields large
sequestering effects, that is, a sensible example of conformal
sequestering with no imposed non-abelian flavor symmetry.%
\footnote{
The above models are governed by the strong dynamics.
We also present a weakly-coupled toy model as a perturbative example
of conformal sequestering in Appendix C.}

After adding mass terms for $H_i$,
\begin{eqnarray}
 W_{mass} = m H_i H_i,
\end{eqnarray}
these models reduce to
the non-calculable SUSY-breaking model discussed above.
The distinguishing feature of these models is that there is no
non-abelian flavor symmetry 
at the fixed point.
Thus, we do not need to impose {\it ad hoc} flavor symmetry in the hidden
sector from the beginning.

{\bf ii) Yukawa couplings}\\
As discussed in the previous section, we can break the unwanted $U(1)$
symmetry associated with the rotation of $H_i$ by adding superpotential
terms. For example, we add the following terms to the
conformally extended non-calculable SUSY-breaking model:
\begin{eqnarray}
W= \sum_{i=1}^{N_{f}} \lambda_i Y_i H_i H_i + \lambda_{ni} Y_i^n,
\end{eqnarray}
together with a deformation 
\begin{equation}
{\mit \Delta} W = \sum_{i=1}^{N_{f}} M_iZ_i(Y_i - m_{i}).  
\end{equation}
Here $Y_i$ and $Z_i$ are additional singlets.
Under the assumption of conformality, the anomalous
dimensions of the fields are given by 
\begin{equation}
\gamma_{Y_{i}} = -2+\frac{6}{n}, \quad \gamma_{10} = 1-\frac{3}{n}, \quad \gamma_{16} = - 11+ \frac{3N_{f}}{2n}. 
\end{equation}
A suitable choice of $N_{f}$ and $n$ might lead to a conformally sequestered model.

More simply, we can introduce a field in a higher representation to
achieve conformality. For example, let us add one symmetric traceless
representation $\Sigma({\bf 54})$ (instead of $N_{f}$ fundamentals)
to the non-calculable SUSY-breaking model. Since
its Dynkin index amounts to $12$, it is equivalent
to $N_{f}=12$ fundamentals in the $\beta$ function of the gauge coupling
\begin{eqnarray}
\beta_{\alpha} = -\alpha^2 \frac{3\times (10-2) -2 (1-\gamma_{16})-12(1-\gamma_{54})}{2\pi-8\alpha}  . \label{nsvz2}
\end{eqnarray}
Under the superpotential
\begin{eqnarray}
W= \lambda_Y Y \Sigma\Sigma + \lambda_n Y^n
\end{eqnarray}
to break the unwanted $U(1)$ symmetry,
we expect to have a SCFT with anomalous dimensions%
\footnote{Corresponding elementary fields
have positive dimensions for $3/2 \le n \le 2$.}
\begin{equation}
\gamma_{Y} = -2+\frac{6}{n}, \quad \gamma_{54} = 1-\frac{3}{n}, \quad \gamma_{16} = - 11+ \frac{18}{n}. 
\end{equation}

If the theory flows to a SCFT,
we obtain a model of conformal sequestering.
Then the sequestering matrix is given by
\begin{eqnarray}
\frac{d}{dt} {\mit \Delta}\ln Z_i = 
- \sum_a\left(\frac{\partial\gamma_i}{\partial\alpha_a}\right)
{\mit \Delta}\alpha_{a} = \sum_kL_{ik} {\mit \Delta}\ln Z_k,
\end{eqnarray}
where $\alpha_\lambda = {|\lambda|^2}/{4\pi}$ and $\alpha_n = {|\lambda_n|^2}/{4\pi}$ with 
\begin{eqnarray}
{\mit \Delta}\alpha_{SO(10)}  &=&
\frac{\alpha_{SO(10)}^{*2}}{2\pi -8 \alpha^*_{SO(10)}}[12{\mit \Delta}\ln Z_{54}
 + 2{\mit \Delta}\ln Z_{16}], \nonumber\\
{\mit \Delta}\alpha_{\lambda}  &=& -\alpha_\lambda^*(2{\mit \Delta}\ln Z_{54}+{\mit \Delta}\ln Z_{Y}), \cr
{\mit \Delta}\alpha_{n}  &=& -\alpha_n^*n{\mit \Delta}\ln Z_{Y}.
\end{eqnarray}
We again note that the sequestering matrix $L_{ik}$ is given by
the Hessian of the renormalization group flow $\partial_{g_k}\beta_i|_*$.
The deformation
\begin{equation}
{\mit \Delta} W =  MZ(Y - m)
\end{equation}
makes low-energy physics
effectively governed by the non-calculable SUSY-breaking model.

\section{Model with no $U(1)$ symmetry enhancement}\label{sec:noSym}

As promised in section 2,
we now try to go beyond the models with no global symmetry
constructed in the previous section.
Namely, we propose to utilize a SUSY-breaking model with no anomalous $U(1)$
symmetry.
Then we encounter no enhanced $U(1)$ symmetry even when we add vector-like 
matter to turn the model into a SCFT, which enables us to present a possibly
simplest model for conformal sequestering.

Our starting point is a calculable variant to the non-calculable
SUSY breaking model of the $SO(10)$ gauge theory
with one {\bf 10} representation $H$ and one {\bf 16} representation $\psi$
\cite{Murayama:1995ng}.
The superpotential is given by
\begin{eqnarray}
W_{tree} = \lambda_\psi \psi \psi H + \frac{1}{2} M H^2,
\label{sup}
\end{eqnarray}
where $M$ denotes a mass scale.
The limit $M \to \infty$ corresponds to the original non-calculable
model with only $\psi$.
Owing to the superpotential (two couplings: $\lambda_\psi$
and $M$ for two multiplets: $\psi$ and $H$),
we have no (anomalous) $U(1)$ symmetry in this SUSY-breaking model.

Now we add a 54-dimensional chiral multiplet $\Sigma$
to make the theory conformal.%
\footnote{We expect that the same model
{\it without} the superpotential also flows to a SCFT
in the IR by calculating 
the conformal $R$ charge through the $a$-maximization procedure 
shown as the last row in table \ref{table1}.
The value is significantly different
from the model {\it with} the superpotential,
so the $U(1)$ breaking seems large as needed. }
Note that we do not require any additional gauge
interactions nor superpotential terms
which would break $U(1)$ rotation of $\Sigma$.
All the $U(1)$ currents are already broken either by
the superpotential terms in Eq.(\ref{sup})
or by the anomaly due to the $SO(10)$ gauge interaction.
The unique possible conformal
$U(1)_R$ assignment is given by
\begin{eqnarray}
R_{10} = 1, \quad R_{16} = \frac{1}{2}, \quad R_{54} = \frac{5}{12},
\end{eqnarray}
which is consistent with the unitarity of the gauge invariant
operators.

If the total theory indeed flows to a nontrivial fixed point,
the conformal sequestering takes place
since the model contains no conserved $U(1)$ charge
(except for $U(1)_R$) even at the fixed point.
After adding a mass term for $\Sigma$, {\it i.e.}
$W_{mass} = m \Sigma\Sigma$, the low-energy
physics is effectively described by
the above SUSY-breaking model.

\section*{Acknowledgements}
We would like to thank M.~Schmaltz for valuable comments on SUSY breaking as
discussed in Appendices.
Y.~N. acknowledges valuable discussion with Y.~Ookouchi on $a$-maximization.
M.~I. and Y.~N. thank the Japan Society for the Promotion of Science
for financial support.

\appendix
\section{SUSY recovery}

In section 3, we have discussed a way to break the unwanted $U(1)$
symmetry by means of superpotential interactions.
For instance, we can add
tree-level Yukawa couplings to break the $U(1)$ symmetry in the
extended IYIT model with $SP(N_c)$ gauge group
\begin{eqnarray}
W_{tree} = \lambda S_{ij} Q^{i}Q^{j}- M_S^2 S_{ij}J^{ij} +\lambda_Y Y\Phi^2 - \lambda_n Y^n + m\Phi^2 ,
\end{eqnarray}
where $J^{ij} = i\sigma_2\otimes {\bf 1}_{N_c+1}$ is the symplectic form.
The second term shifts the origin of the meson superfield $M^{ij} =
Q^{i}Q^{j}$ and serves as a regularization as we will see. 

For $m=M_S=0$, the theory possibly flows to a nontrivial
conformal fixed point for
$g_{SP(3)}, \lambda$, $\lambda_Y$, and $\lambda_n$.
Since all the interactions are strong for a suitable choice of $n$, the
sequestering matrix is expected to be order one. 
Unfortunately, however, SUSY is not broken in this example.
In this appendix, we confirm that the theory recovers SUSY as the
existence of Higgs vacua.

To see this, we utilize the glueball superpotential
technique (see Ref.~\cite{Brandhuber:2003va,Nakayama:2003ri}).
First of all,
let us assume a supersymmetric vacuum and write down the Konishi anomaly
equation~\cite{Konishi}
\begin{eqnarray}
\bar{D}^2J = \phi_j\frac{\partial W_{tree}}{\partial \phi_i} + \frac{T(r_i)}{32\pi^2}\mathrm{Tr}{\cal W}^2\delta_{ij} ,
\end{eqnarray}
where $T(r_i)$ is the Dynkin index of $i$-th matter,
 evaluated at such a hypothetical vacuum:
\begin{eqnarray}
2\lambda\langle S_{ij} M^{kj}\rangle &=& \delta^k_i S, \cr
\lambda \langle S_{ij} M^{kl}\rangle -M_S^2J^{kl}\langle S_{ij} \rangle &=& 0, \cr
m\langle \Phi^2 \rangle + \lambda_Y \langle Y\Phi^2\rangle &=& S, \cr
\lambda_Y\langle \Phi^2\rangle - n \lambda_n \langle Y^{n-1}\rangle &=& 0  ,
\end{eqnarray}
where $S = -\frac{1}{32\pi^2}\langle \mathrm{Tr} {\cal W}^2 \rangle$ is the so-called glueball
superfield. When $M_S$ is 0, the supersymmetric vacuum exists
if and only if $S=0$.

For nonzero $M_S$, we can solve these
equations to derive the vacuum expectation values of matter
superfields in terms of $S$.
The solutions have two branches, one of which corresponds to the classical unHiggsed branch:
\begin{eqnarray}
\langle M^{ij}\rangle &=& \frac{M_S^2}{\lambda} J^{ij}, \cr
\langle S_{ij}\rangle &=& \frac{1}{2}\frac{S}{M_S^2} ({J^{-1}})_{ij}, \cr
\langle \Phi^2\rangle &=& \frac{S}{m} + \mathcal{O}(S^2), \cr
\langle Y^{n-1} \rangle &=& {\frac{\lambda_YS}{n\lambda_nm}} + \mathcal{O}(S^{2})  .
\end{eqnarray}
The other one corresponds to the Higgsed branch:
\begin{eqnarray}
\langle M^{ij}\rangle &=& \frac{M_S^2}{\lambda} J^{ij}, \cr
\langle S_{ij}\rangle &=& \frac{1}{2}\frac{S}{M_S^2} (J^{-1})_{ij}, \cr
\langle \Phi^2\rangle &=& \frac{n\lambda_n}{\lambda_Y}\left(-\frac{m}{\lambda_Y}\right)^{n-1}-\frac{n-1}{m}S  + \mathcal{O}(S^2), \cr
\langle Y \rangle &=& -\frac{m}{\lambda_Y}+\frac{1}{n\lambda_n}\left(-\frac{\lambda_Y}{m}\right)^{n-1}S+ \mathcal{O}(S^2)  .
\end{eqnarray}

Now we integrate each set
of equations to obtain the effective glueball
superpotential up to an integration constant $C(S)$.

In the unHiggsed branch, we obtain
\begin{eqnarray}
W_{eff}(S) = (N_c+1)S\ln \frac{\Lambda^2 \lambda}{M_S^2} + S\ln \frac{m}{\Lambda}+ \mathcal{O}(S^{n \over n-1}) ,
\end{eqnarray}
where $\Lambda$ denotes the dynamical scale of the gauge interaction and 
$C(S)$ is determined from the method explained in Ref.~\cite{Nakayama:2003ri}. 
The superpotential is singular
in the $M_S\to 0$ limit unless $S=0$. 
However, for $S=0$, the $F$-term condition is not satisfied to imply
broken SUSY.
This is consistent with the fact that the low-energy dynamics in this branch
are nothing but those of the IYIT model.

On the other hand,
in the Higgsed branch, the effective glueball superpotential takes a form
\begin{eqnarray}
W_{eff}(S) = (N_c+1)S\ln \frac{\Lambda^2 \lambda}{M_S^2} -\lambda_n\left(-\frac{m}{\lambda_Y}\right)^{n} + S\left(\ln \frac{\lambda_Y^n S}{m^{n-1} \lambda_n\Lambda} - 1\right) + \mathcal{O}(S^{2})  .
\end{eqnarray}
The extremization condition of the glueball superpotential is given by
\begin{eqnarray}
\frac{\lambda_Y^n S}{\lambda_nm^{n-1}\Lambda} = \left(\frac{M_S^2}{\Lambda^2\lambda}\right)^{N_c+1} \ ,
\end{eqnarray}
and this gives a solution $S\to0$ as $M_S \to 0$.
Therefore, there exists a supersymmetric vacuum for this branch.
This means that the model shows a recovery of SUSY due to the $U(1)$
breaking interaction.

We can use a different relevant deformation to obtain a SUSY-breaking
model in IR, avoiding problematic Higgs vacua.
Namely, instead of the mass term $m \Phi^2$,
we introduce the following superpotential terms:
\begin{eqnarray}
 {\mit \Delta} W = MZ(Y - m).
\end{eqnarray}
Similarly the effective glueball superpotential is given by
\begin{eqnarray}
W_{eff}(S) = (N_c+1)S\ln \frac{\Lambda^2 \lambda}{M_S^2}
-\lambda_n{m}^{n} + S\ln \frac{\lambda_Ym}{\Lambda}
\end{eqnarray}
to imply dynamically broken SUSY.%
\footnote{In the broken SUSY case,
the effective glueball superpotential here does not necessarily yield
a good low-energy description of the model.
An attempt to construct an effective action in a similar SUSY-breaking
model can be found in Ref.~\cite{Konishi:1986qm}.}

\section{From gauge to Yukawa}

In section 3, we have discussed two possible ways
to strongly break unwanted $U(1)$ 
symmetry which hinders the conformal sequestering: one is to break it 
anomalously by gauge interaction and the other is to break it explicitly by 
superpotential interaction.
The latter Yukawa-type deformation
possibly has an intimate relationship with the former gauge-type one.
In this Appendix, we give an example to indicate such an interplay.

\subsection{effective theory}

Let us consider the following example. The extended IYIT model with 
$SP(3)$ has an unwanted $U(1)$ symmetry which rotates $S_{ij}$, $Q^i$, and 
$Q'^j$ simultaneously. To attain the conformal sequestering, we have 
introduced $SP(1) \times SP(1)$ gauge symmetry which breaks the $U(1)$ 
rotation of $Q'^j$ by anomaly of $SP(1) \times SP(1)$ gauge interaction
in Ref.~\cite{Ibe:2005pj}.
The new theory with the $SP(1) \times SP(1)$ gauging also flows 
to a conformal field theory in the IR, ensuring the conformal sequestering 
of the gauged theory.

Suppose $g_{{}_{SP(3)}} \ll g_{{}_{SP(1)}}$
at an intermediate scale between 
$\Lambda_{UV}$ and $\Lambda_{CFT}$.
Then we can solve the strong dynamics of 
the $SP(1) \times SP(1)$ theory first and consider an effective $SP(3)$ 
gauge theory. This effective $SP(3)$ gauge theory has the the following 
matter contents: 8 fundamental representations $Q^i$, 28 singlets
$S^{ij}$, 2 antisymmetric ($14$-dimensional) representations $A_{ab}$
and 2 additional singlets $B$. 
The $A_{ab}$ and $B$ are composite meson superfields of the $Q'$ 
in the strong $SP(1) \times SP(1)$ dynamics:
\begin{eqnarray}
 Q'_aQ'_b \sim A_{ab} + J_{ab} B,
\label{composition}
\end{eqnarray}
where $J_{ab}$ denotes the symplectic form.

The induced effective $SP(3)$ gauge theory
has the following effective superpotential:
\begin{eqnarray}
W_{eff} &=& \lambda_1\left({A^{(1)}_{ab}A^{(1)}_{cd}A^{(1)}_{ef}}\epsilon^{abcdef} + 
3B^{(1)}{A^{(1)}_{ab}A^{(1)}_{cd}J_{ef}}\epsilon^{abcdef} + 24 (B^{(1)})^3 \right) \cr
& &+ 
\lambda_2\left({A^{(2)}_{ab}A^{(2)}_{cd}A^{(2)}_{ef}}\epsilon^{abcdef} + 
3B^{(2)}{A^{(2)}_{ab}A^{(2)}_{cd}J_{ef}}\epsilon^{abcdef} + 24 (B^{(2)})^3 \right).
\label{gtoY}
\end{eqnarray}
This effective potential comes from the effective superpotential of
the $SP(1)$ gauge theory such as $\mathrm{Pf}(Q'_aQ'_b)$.
The $SP(3)$ dynamics together with this superpotential lead to the 
same IR dynamics as those of the $SP(3) \times SP(1)^2$ gauge theory.
The unwanted $U(1)$ symmetry
which would be given by a rotation of $A$ and $B$
is broken explicitly by the superpotential terms. 

Let us perform a simple consistency check: if $W_{eff}$ possesses a
nontrivial IR fixed point, the conformal $R$ charge of $A$ and $B$ is
given by 
\begin{eqnarray}
R_{A} = R_{B} = \frac{2}{3},
\end{eqnarray}
which indeed agrees with the direct $SP(3) \times SP(1) \times SP(1)$
computation: $\gamma_{Q'} = -1$ in terms of
Eq.(\ref{composition}). The vanishing of the $\beta$ function for the gauge coupling then shows that $\gamma_{Q} = -1$ also in agreement with the $SP(3) \times SP(1) \times SP(1)$ results.

\subsection{SUSY breaking}

We have seen that the anomalous breaking of the $U(1)$
symmetry by gauging and the explicit breaking by the superpotential are
dual descriptions of the same physics in the above example.
One potential subtlety is
that the phase of SUSY after adding a mass term for $Q'$, or
equivalently the addition of tadpole $m^2B$ in the superpotential.
From the former
viewpoint, dynamical SUSY breaking of the model is obvious due to
the decoupling argument. However it could appear subtle in the dual
Yukawa viewpoint.
In this subsection, we explain SUSY breaking in the composite model
with a particular Yukawa interaction.

We consider the $SP(3)$ gauge theory with IYIT
matter sector coupled to an extra antisymmetric-tensor ({\bf 14})
representation $A_{ab}$ and a singlet $B$.
The superpotential generically takes a form
\begin{eqnarray}
W = \lambda S_{ij} Q^{i}Q^{j} -  M_S^2 S_{ij}J^{ij} - m^2B 
+\alpha B^3 + \beta B A^{}_{ab}A^{}_{cd}J_{ef}\epsilon^{abcdef} +  \gamma A^{}_{ab}A^{}_{cd}A^{}_{ef}\epsilon^{abcdef}.
 \label{super}
\end{eqnarray}
The second term was introduced as a regularization as in Appendix A.
In the following, we argue that when $\alpha,\beta,\gamma$ satisfy a
specific relation $\alpha = \frac{\beta}{3} = \frac{\gamma}{24} = c$ inferred from
Eq.(\ref{gtoY}),
this model causes dynamical SUSY breaking.%
\footnote{The condition of SUSY breaking
is actually equivalent to the condition that this model can be
derived by integrating out the strongly coupled $SP(1)$ gauge
interaction as in the previous subsection.}

Let us show this by means of the glueball superpotential
technique as in Appendix A.
First we assume a supersymmetric vacuum and write down the Konishi anomaly
equation evaluated at such a hypothetical vacuum:
\begin{eqnarray}
2\lambda\langle S_{ij} M^{kj}\rangle &=& \delta^k_i S \cr
\lambda \langle S_{ij} M^{kl}\rangle -M_S^2J^{kl}\langle S_{ij} \rangle &=& 0 \cr
2 \langle B^{}A^{}_{ab}A^{}_{cd}J_{ef}\epsilon^{abcdef} \rangle+\langle A^{}_{ab}A^{}_{cd}A^{}_{ef}\epsilon^{abcdef}\rangle &=& \frac{4S}{3c} \cr
\langle 24 B^2\rangle +3 \langle A^{}_{ab}A^{}_{cd}J_{ef}\epsilon^{abcdef}  \rangle &=& \frac{m^2}{3c}.
 \label{konishi2}
\end{eqnarray}
We note that the supersymmetric vacuum exists
if and only if $S=0$ for $M_S=0$.
For $M_S \neq 0$, we can solve
Eq.(\ref{konishi2}) perturbatively around $S=0$,
which results in
\begin{eqnarray}
\langle M^{ij}\rangle &=& \frac{M_S^2}{\lambda} J^{ij} \cr
\langle S_{ij}\rangle &=& \frac{S}{2M_S^2} (J^{-1})_{ij} \cr
\langle A^{}_{ab}A^{}_{cd}J_{ef}\epsilon^{abcdef} \rangle &=& {\cal O}(\frac{S}{\sqrt{cm^2}}) \cr
\langle B \rangle &=& \pm\sqrt{\frac{m^2}{72c}} + {\cal O}(S) .
\end{eqnarray}
Actually, Eq.~(\ref{konishi2}) have several solutions, two of which
survive in the $S\to0$ limit. The other solutions run away to infinity
of $S_{ij}$
for $M_S\to 0$.%
\footnote{If we adopted a general cubic potential
Eq.(\ref{super})
instead of the choice
in Eq.(\ref{konishi2}),
the other solutions would stay in the classical Higgs branch
and eventually lead to the recovery of SUSY.}

Now we integrate the above set of
equations to obtain the effective glueball
superpotential to the leading order in $S$:
\begin{eqnarray}
W_{eff}(S) = \mp \frac{m^{3}}{9\sqrt{2C}}
+ c_1 S + (N_c+1)S\ln \frac{\Lambda^2 \lambda}{M_S^2}  + {\cal O}(S^2),
 \label{effec}
\end{eqnarray}
where $c_1 = \ln({m^{12}}/{c^{4}\Lambda^{12}})$.
Owing to the $S\ln (\lambda\Lambda^2 /M_S^2)$ term,
the supersymmetric vacuum would be
given by $S=0$ in the $M_S\to 0$ limit if any.
However, then, the $F$-term
condition would not be satisfied in Eq.~(\ref{effec})
to imply dynamical SUSY breaking in this model.

\section{Perturbative example}

In section 3, we have considered nonperturbative models
for conformal sequestering.
In this Appendix, we provide
a perturbative example of the $SO(6)\times SO(6) \times
SO(6)$ gauging of the conformally extended $SO(10)$ non-calculable
models with bi-fundamentals transforming as $({\bf 10}, {\bf 6}_i)$ for
$i=1,2,3$.
The conformal $R$ charges are determined to be
$R_{10} = R_{16} = \frac{3}{5}$, which is close to the free field value $R=\frac{2}{3}$ and suggests a possibility of perturbative computation. 

At the one-loop level, the anomalous dimension
of a matter indexed by $i$ in a gauge theory is given by a formula
\begin{eqnarray}
\gamma_i(\alpha) = -\frac{\alpha}{\pi} C_2(r_i) +\mathcal{O}(\alpha^2);
\quad
C_2(r) = \frac{|G|}{|r|}T(r),
\end{eqnarray}
where $|G|$ and $|r|$ denote
the dimensions of the group and the representation,
respectively, and $T(r)$ is the Dynkin index.%
\footnote{For the gauge group $SO(2N)$, $T({\rm fundamental})=1$ and
$T({\rm spinor})=2^{N-4}$.}

In this model, one-loop calculation shows 
\begin{eqnarray}
\gamma_{10} &=& -\frac{9}{2\pi}\alpha_{SO(10)} - \frac{5}{2\pi} \alpha_{SO(6)}, \nonumber\\
\gamma_{16} &=& -\frac{45}{8\pi} \alpha_{SO(10)} .
\end{eqnarray}
Then, the vanishing of the $\beta$ functions
\begin{eqnarray}
\beta_{\alpha_{SO(10)}} &=& -\alpha_{SO(10)}^2 \frac{3\times (10-2) -2 (1-\gamma_{16})-18(1-\gamma_{10})}{2\pi-8\alpha_{SO(10)}} , \cr
\beta_{\alpha_{SO(6)}} &=& -\alpha_{SO(6)}^2 \frac{3\times (6-2) -10(1-\gamma_{10})}{2\pi-4\alpha_{SO(6)}} \
\end{eqnarray}
determines $\alpha_{SO(10)}^* = {8\pi}/{225}$ and $\alpha_{SO(6)}^* = {2\pi}/{125}$, which suggests that the one-loop approximation is not so bad in the sense of Ref.~\cite{BZ}. 

The sequestering matrix is given by
the Hessian $\partial_{\alpha_i} \beta_k|_{g=g^*}$ \cite{Ibe:2005pj}.
Its eigenvalues are evaluated as $(0.0004,0.07)$, and its smallest value
amounts to the efficiency of the conformal sequestering.
Although we see that the
perturbatively obtained conformal sequestering is too small for
phenomenological
applications, we have presented one example for definiteness.


\begin{thebibliography}{99}
\bibitem{BZ}
  T.~Banks and A.~Zaks,
  Nucl.~Phys.~{\bf B196} (1982) 189.
\bibitem{SC}
  M.~Flato and C.~Fronsdal,
  Lett.~Math.~Phys.~{\bf 8} (1984) 159; \\
  V.K.~Dobrev and V.B.~Petkova,
  Phys.~Lett.~{\bf B162} (1985) 127.
\bibitem{CGT}
  N.~Seiberg, arXiv:hep-th/9411149.
\bibitem{LS} 
  M.~Luty and R.~Sundrum,
  arXiv:hep-th/0105137; arXiv:hep-th/0111231.
\bibitem{Dine}
  M.~Dine, P.J.~Fox, E.~Gorbatov, Y.~Shadmi, Y.~Shirman and S.~Thomas,
  arXiv:hep-ph/0405159.
\bibitem{Sundrum}
  R.~Sundrum, arXiv:hep-th/0406012.
\bibitem{NS} 
  A.E.~Nelson and M.J.~Strassler,
  arXiv:hep-ph/0006251; arXiv:hep-ph/0104051; \\
  T.~Kobayashi and H.~Terao,
  arXiv:hep-ph/0103028; \\
  T.~Kobayashi, H.~Nakano and H.~Terao,
  arXiv:hep-ph/0107030; \\
  T.~Kobayashi, H.~Nakano, T.~Noguchi and H.~Terao,
  arXiv:hep-ph/0202023.
\bibitem{Ibe:2005pj}
  M.~Ibe, K.~I.~Izawa, Y.~Nakayama, Y.~Shinbara and T.~Yanagida,
  arXiv:hep-ph/0506023.
\bibitem{IYIT} 
  Izawa~K.-I.~and T.~Yanagida, arXiv:hep-th/9602180; \\
  K.A.~Intriligator and S.~Thomas, arXiv:hep-th/9603158.
\bibitem{Affleck:1984mf}
  I.~Affleck, M.~Dine and N.~Seiberg,
  Phys.\ Lett.\ B {\bf 140}, 59 (1984).
\bibitem{Murayama:1995ng}
  H.~Murayama, arXiv:hep-th/9505082.
\bibitem{NSVZ}
  V.A.~Novikov, M.A.~Shifman, A.I.~Vainshtein and V.I.~Zakharov,
  Nucl.~Phys.~{\bf B229} (1983) 381; \\
  M.A.~Shifman and A.I.~Vainshtein,
  Nucl.~Phys.~{\bf B277} (1986) 456;
  Nucl.~Phys.~{\bf B359} (1991) 571; \\
  N.~Arkani-Hamed and H.~Murayama, arXiv:hep-th/9707133.
\bibitem{Anselmi:1996mq}
  D.~Anselmi, M.~T.~Grisaru and A.~Johansen,
  arXiv:hep-th/9601023.
\bibitem{Intriligator:2003jj}
  K.~Intriligator and B.~Wecht,
  arXiv:hep-th/0304128; \\
  E.~Barnes, K.~Intriligator, B.~Wecht and J.~Wright,
  arXiv:hep-th/0408156.
\bibitem{ISS}
  K.A.~Intriligator, N.~Seiberg and S.~H.~Shenker,
  arXiv:hep-ph/9410203; \\
  See also K.~Intriligator, arXiv:hep-th/0509085.
\bibitem{Pouliot:1996zh}
  P.~Pouliot and M.~J.~Strassler, arXiv:hep-th/9602031; \\
  T.~Kawano, arXiv:hep-th/9602035.
\bibitem{Ans}
  D.~Anselmi, J.~Erlich, D.Z.~Freedman, and A.A.~Johansen,
  arXiv:hep-th/9711035.
\bibitem{private}
  T.~Kawano, Y.~Ookouchi, Y.~Tachikawa, and F.~Yagi,
private communication.
\bibitem{Brandhuber:2003va}
  A.~Brandhuber, H.~Ita, H.~Nieder, Y.~Oz and C.~Romelsberger,
  arXiv:hep-th/0303001.
\bibitem{Nakayama:2003ri}
  Y.~Nakayama, arXiv:hep-th/0306007.
\bibitem{Konishi}
  K.~Konishi, Phys.\ Lett.\ {\bf B135} (1984) 439; \\
  K.~Konishi and K.~Shizuya,
  Nuovo Cim.\ {\bf A90} (1985) 111.
\bibitem{Konishi:1986qm}
  K.~Konishi and G.~Veneziano,
  Phys.\ Lett.\ {\bf B187} (1987) 106.
\end{thebibliography}
\end{document}